\documentclass[fleqn,usenatbib]{mnras}

\usepackage{newtxtext,newtxmath}

\usepackage[T1]{fontenc}
\usepackage{ae,aecompl}

\usepackage{graphicx}	
\usepackage{amsmath}	
\usepackage{amssymb}	
\usepackage{color}
\usepackage[usenames,dvipsnames,svgnames,table]{xcolor}
\hypersetup{
    citecolor = {gray},
    linkcolor = {black}
}

\newcommand{\bn}{\begin{enumerate}}
\newcommand{\en}{\end{enumerate}}
\newcommand{\bi}{\begin{itemize}}
\newcommand{\ei}{\end{itemize}}

\newcommand{\Zsun}{Z_\odot}

\newcommand{\Msun}{M_\odot}

\newcommand{\buv}{\beta_{\rm UV}}
\newcommand{\buvpiii}{\beta_{\rm UV,PopIII}}

\newcommand{\Myr}{\rm Myr}

\newcommand{\Ebv}{E(B-V)}

\newcommand{\angstrom}{\textup{\AA}}

\title{Dust extinction in the first galaxies}

\author[Jaacks et al.]{Jason Jaacks$^1$\thanks{email: jaacks@astro.as.utexas.edu}, Steven L. Finkelstein$^1$ and Volker Bromm$^1$ \vspace{0.3cm}\\
$^1$ Department of Astronomy, The University of Texas at Austin, Austin, TX 78712\\
}

\date{Accepted XXX. Received YYY; in original form ZZZ}

\pubyear{2017}
\begin{document}
\label{firstpage}
\pagerange{\pageref{firstpage}--\pageref{lastpage}}
\maketitle

\begin{abstract}
Using cosmological volume simulations and a custom built sub-grid model for Pop~III star formation, we examine the baseline dust extinction in the first galaxies due to Pop~III metal enrichment in the first billion years of cosmic history. We find that while the most enriched, high-density lines of sight in primordial galaxies can experience a measurable amount of extinction from Pop~III dust ($\Ebv_{\rm max}=0.07,\ A_{\rm V,max}\approx0.28$), the average extinction is very low with $\left< \Ebv \right> \lesssim 10^{-3}$. We derive a power-law relationship between dark matter halo mass and extinction of $\Ebv\propto M_{\rm halo}^{0.80}$. Performing a Monte Carlo parameter study, we establish the baseline reddening of the UV spectra of dwarf galaxies at high redshift due to Pop~III enrichment only. With this method, we find $\left<\buv\right>-2.51\pm0.07$, which is both nearly halo mass and redshift independent.
\end{abstract}




\section{Introduction} 
\label{sec:intro}

One of the fundamental goals in the study of galaxy evolution is the quest to observe the first galaxies. With the launch of the {\it James Webb Space Telescope (JWST)} on the horizon, this goal may be attainable. However, challenging questions remain. {\it How will we know when we have found them, and what will be their observational signatures?} To begin, we must have a workable definition for what a ``first galaxy'' is. For the purpose of this work, where we straddle the line between theory and observations, we define a ``first galaxy'' to be the earliest observable object which is gravitationally bound to a dark matter halo, and which is undergoing, or has undergone, Population III (Pop~III) and at least one episode of Population II (Pop~II) star formation \citep{Bromm:11}. Since we are requiring Pop~II stars to be present, we limit our analysis to dark matter halos with sufficient mass to promote efficient cooling via \ion{H}{i} ($M_{\rm halo} \gtrsim 10^7\ \Msun$).

Within this definition one would expect these galaxies to host young stellar populations and contain gas which is minimally enriched with metals from supernova (SN) events.  Observationally, systems with the above characteristics would exhibit very blue, steep UV power-law spectral slopes ($f_\lambda\propto\lambda^\beta$), as young, hot stars dominate the spectra. Intrinsic UV stellar spectral slopes of $\buv\sim-3.0$ are predicted from stellar evolution models for young, very low metallicity ($Z_*\lesssim 10^{-3}Z_{\odot}$) stellar populations \citep{Schaerer:03b}. Any deviations from this intrinsic slope are typically attributed to physical properties of the interstellar medium (ISM) through which the photons propagate, such as metallicity, density, dust content, and ionization state. 

Observations at intermediate redshifts ($z\sim3$) have shown that $\buv$ is correlated with far-infrared (FIR) dust emission \citep[e.g.][]{Meurer:99,Reddy:12b}. These results indicate that UV photons are readily absorbed and scattered by dust grains which are in turn heated, re-radiating in the FIR. Intense efforts extend to higher redshifts ($z\geq6$), as $\buv$ is a key parameter to model reionization. It can be directly measured from broadband photometry, and thus is accessible out to the highest redshifts currently reached ($z \sim$ 10; \citealt{wilkins:16}). Early results from \cite{Bouwens:10d} and \citet{Finkelstein:10} found evidence for $\buv\sim-3.0$ ($\pm$0.2--0.5), suggesting extremely metal-poor stellar populations and no dust extinction, albeit with no conclusive evidence for primordial star formation. However, more recent studies have benefited from larger sample sizes and improved bias corrections, finding $\left<\buv\right>\sim -2.2$ to $-2.4$ ($\pm$ 0.30) for faint ($M_{UV} \sim -$18) galaxies at $z\simeq7$, again indicative of little dust attenuation, but non-primordial stellar populations \citep{McLure:11,Dunlop:12,Finkelstein:12c,Bouwens:14a}. These results imply that galaxies hosting primordial star-formation must reside at even higher redshifts, and/or at fainter luminosities.

There have been a number of pioneering studies which have utilized sophisticated spectral synthesis and evolution codes to predict the photometric properties of the first galaxies \citep[e.g.][]{Schaerer:02,Zackrisson:13}. While providing detailed predictions for the spectral energy distribution, given assumptions regarding key physical properties (such as halo mass, escape fraction, and age), these investigations lack the ab initio cosmological context that numerical simulations can provide. The latter enable to trace the realistic transport of heavy chemical elements in the evolving three-dimensional cosmic web, as is done here. There has also been a vigorous effort to leverage cosmological simulations in conjunction with spectral evolution codes to predict photometric ``first galaxy'' properties \citep[e.g.][]{wilkins:16,Barrow:17,Zackrisson:17}.  However, these works focus on the properties of typical, already more evolved, high-redshift galaxies. Here, on the other hand, our focus is on the chemically most primitive galaxies, where the first Pop~II clusters form out of material that has only been enriched by Pop~III \citep{Jaacks:17a}. This allows us to establish a baseline beta slope, the extreme blue limit of $\beta$ that may be detectable with the {\it JWST}. Our goal for this work is to help guide interpretations of the next generation deep-field surveys, in particular in terms of how close to ``first light'' a given source is.

 The paper is structured as follows. In Section~\ref{sec:methods} we describe our numerical methodology, and in Section~\ref{sec:results} and \ref{sec:sum} we present our results and conclusions.   

\section{Numerical methods} 
\label{sec:methods}


For this work, we utilize a customized version of the publicly available next generation hydrodynamics code {\small GIZMO}, which employs a Lagrangian meshless finite-mass (MFM) methodology for solving the equations of fluid dynamics (for details regarding simulations and sub-grid models, see \citealt{Jaacks:17a}). {\small GIZMO} offers improved numerical accuracy and efficiency by combining features of smoothed particle hydrodynamics (SPH) and adaptive mesh refinement (AMR) codes.

\subsection{Simulations}
\label{subsec:sims}
Our simulation volume, designed to approximately replicate a single pointing with {\it JWST} at redshift $z\sim 10$, has a box size of $4h^{-1}$ Mpc and contains $512^3$ particles of both gas and dark matter.  
We adopt a $\Lambda$ cold dark matter ($\Lambda$CDM) cosmology, consistent with the recent {\it Planck} results: $\Omega_m=0.315$, $\Omega_\Lambda=0.685$, $\Omega_b=0.047$, $\sigma_8=0.829$, and $H_0=100h\,{\rm km\ s^{-1}\ Mpc^{-1}}=67.74\ {\rm km\ s^{-1}\ Mpc^{-1}}$ \citep{Planck:16}.   Our initial conditions are generated at $z=250$, using the {\small MUSIC} initial conditions generator \citep{Hahn:11}.  

The simulations employ our custom-built Pop~III legacy (P3L) star formation sub-grid model which focuses on the long-term impact of Pop~III on the surrounding medium. Specifically, our model is designed to track metals which enrich the early ISM through SN explosions. The P3L approach allows each Pop~III star forming region to have a randomly selected population, drawn from a given initial mass function (IMF), here taken to be top heavy with a slope of $\alpha=-0.17$ and an exponential cutoff below $M_*=20\ \Msun$ ( see Section~\ref{sec:sum} for detailed motivation). This random process endows each region with a unique metal enrichment, in terms of amount and spatial extent. We also include a Pop~II proxy (P2P) sub-grid model to approximate the ionization and Lyman-Werner feedback from contemporaneous Pop~II star formation, but neglect any additional metal enrichment from Pop~II.

Dark matter haloes are identified with a post processing 3D friends-of-friends (FOF) algorithm, using a minimum particle number of $32$ and a linking length of $0.15$ times the inter-particle distance.  Gas particles and their respective properties are then associated with each halo by searching within its virial radius. Grouping and data extraction are aided by the {\it yt} \citep{Turk:yt} and {\it Ceasar} \citep{pygr} software packages.

\subsection{Interstellar extinction}
\label{subsec:ext}
Determining the dust extinction in the high-$z$ ISM is non-trivial, both observationally and theoretically.  The primary difficulty lies in ascertaining the quantity and physical nature of the dust particles, including their size, composition, shape, and optical properties. These details are folded into an empirically determined extinction curve \citep[e.g.,][]{Calzetti.etal:00}, where the optical depth is
\begin{equation}
\tau_{\lambda} = 0.921 k(\lambda) E(B-V).
\label{eq:tauebv}
\end{equation}
The wavelength dependence of the Calzetti et al.\ attenuation curve is expressed as
\begin{equation}
k(\lambda) = 
\begin{cases}
2.659 (-1.857 + 1.040/\lambda) + 4.05,\\
{\rm for\ } 0.63 \mu m \leq \lambda \leq 2.20 \mu m \\
2.659 (-2.156 + 1.509/\lambda - 0.198/\lambda^2 + 0.011/\lambda^3) + 4.05, \\
{\rm for\ }0.12 \mu m \leq \lambda \leq 63 \mu m.
\end{cases}
\end{equation}
Assuming an intrinsic model spectrum, the overall $E(B-V)$ can then be determined by fitting to the observed broadband photometry of the system. The exact details regarding the physical properties of dust in the early ISM are highly uncertain. Therefore, we will assume for simplicity that dust from Pop~III star formation is characterized by the Calzetti law. Solving the radiative transfer equation, simplified for pure absorption, we have
\begin{equation}
f_\lambda = f_{\lambda}(0)e^{-\tau_\lambda}.
\label{eq:rt}
\end{equation}
Here, $f_{\lambda}(0)$ is the intrinsic flux representing the stellar population (see Section~\ref{subsec:ssp}), and $\tau_\lambda$ is the total optical depth along the line of sight (LOS), given by Equ.~\ref{eq:tauebv}.

From our simulations, we can determine the dust reddening for a given system with the normalized relation \citep{Bohlin:78,Rachford:09}
\begin{equation}
\Ebv \simeq \frac{N_{\rm HI}}{5.8\times10^{21} {\rm cm^{-2}}}\frac{Z}{Z_\odot}\frac{\zeta}{\zeta_\odot}
\mbox{\ .}
\label{eq:ebv}
\end{equation}
We estimate the neutral hydrogen column via $N_{\rm HI}\simeq\rho_{\rm gas}/(\mu m_{\rm H})L_{\rm char}$, where $L_{\rm char}$ is the characteristic length scale of the system.  The metallicity ($Z$) is extracted directly from the enriched gas, and we set the metal-to-dust ratio to $\zeta/\zeta_\odot\equiv1.0$, where $\zeta_\odot\approx0.50$.

\subsection{Stellar populations}
\label{subsec:ssp}
In this work, we utilize Pop~II stellar clusters as flashlights which illuminate the Pop~III enriched ISM of our simulated galaxies. To represent a Pop~II intrinsic simple stellar population (SSP), we adopt models from \citet{Schaerer:03b}, containing evolutionary tracks for very low metallicity ($Z_*\lesssim 10^{-2}\ \Zsun$). This value is consistent with the maximum Pop~III enrichment found in \citet{Jaacks:17a}, and represents the ISM conditions from which Pop~II stars form. We consider young stellar populations with ages of $10\ \Myr$ and $50\ \Myr$ which have experienced a constant star formation history, with a \cite{Salpeter:55} IMF over a mass range of $1-100\ \Msun$. Note that we intentionally limit the age of our stellar populations in order to represent the extreme scenario studied here, where Pop~II star formation is just beginning, and spectra exhibit their bluest shapes.

The resulting intrinsic spectra are shown by the solid black and gray lines in Figure~\ref{fig:fig1}, where the flux is normalized to a Pop~II stellar mass of $1\ \Msun$, which can be scaled to the total stellar mass of any galaxy. Since nebular continuum emission is a potential reddening factor, we show  $\buv$ assuming Lyman continuum escape fractions of $f_{\rm esc}=1$ and $0$.

\section{Results} 
\label{sec:results}
\subsection{Mean halo extinction}
\label{subsec:halo}

\begin{figure}
\begin{center}
\includegraphics[scale=0.34] {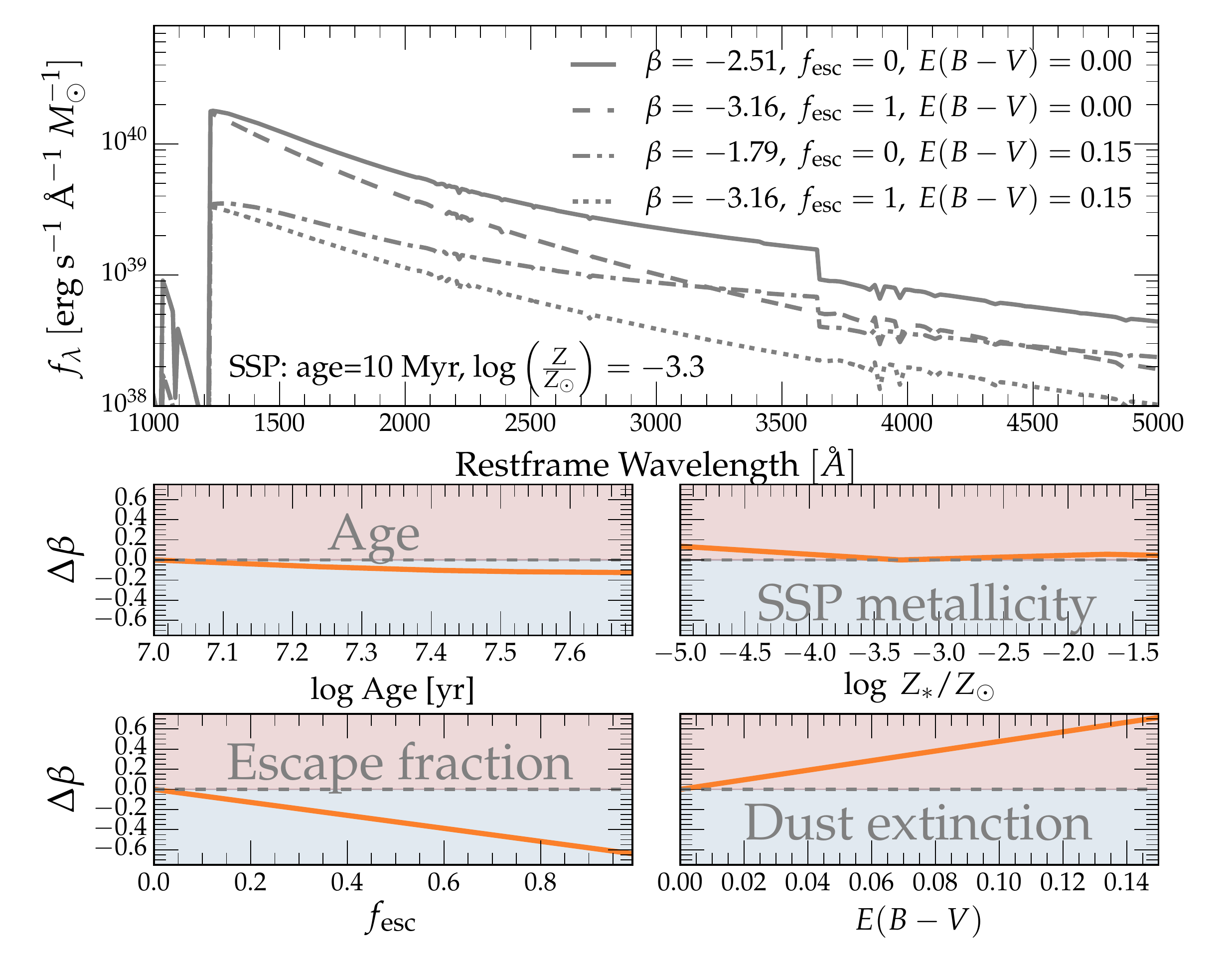}
\caption{Parameter sensitivity of the UV spectral slope $\buv$.  The series of gray lines represents an SSP with and age of 10 Myr and $\log(Z/\Zsun)=-3.3$, with varying values of $f_{\rm esc}$ and \Ebv. $\buv$ are measured by performing a least-squares fit over a rest-frame wavelength range of $1200-2600\ \angstrom$.  The bottom panels show the impact of individual parameters (age, metallicity, $f_{\rm esc}$, and $\Ebv$) on $\buv$ where $\Delta\beta\equiv \beta_{\rm SSP}-\beta_{\rm param}$.  For each panel only the indicated variable is varied and all others are fixed. It should be noted that, while the limited age range considered here has little impact on $\Delta \beta$, stellar populations older than those studied here would be characterized by significantly redder values of $\beta$.
}
\label{fig:fig1}
\end{center}
\end{figure}

We first explore the mean Pop~III extinction for each halo identified in our simulations at $z$=7.5.  The extinction value for each halo is calculated by assuming the total gas and metals are distributed homogeneously within the half mass radius ($R_{\rm 1/2}$) of the halo.  These halo averaged properties are then employed to calculate the halo averaged Pop~III extinction, using Equ.~\ref{eq:ebv} and $L_{\rm char} = R_{\rm 1/2}$. 

\begin{figure}
\begin{center}
\includegraphics[scale=0.34] {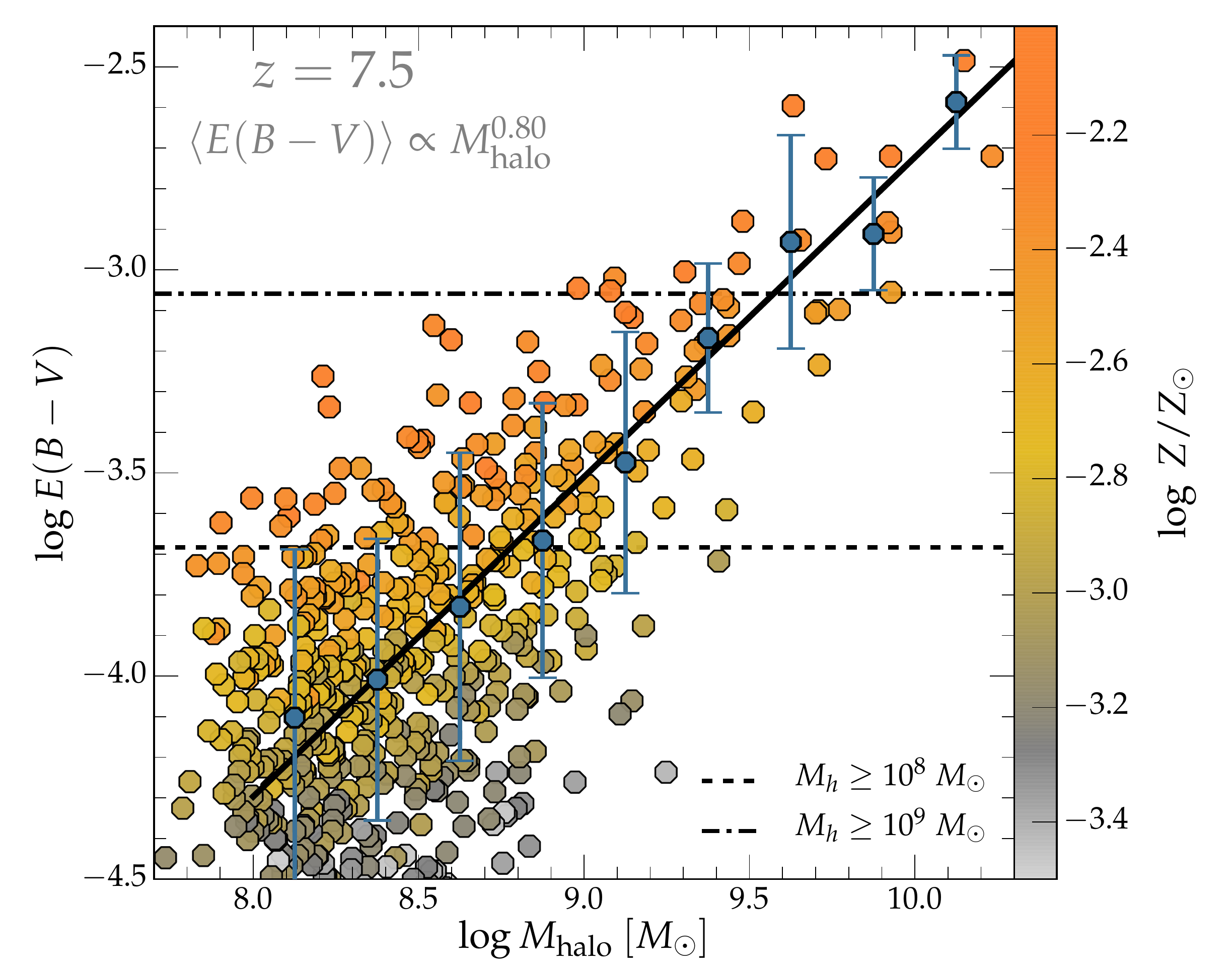}
\caption{Dust reddening for haloes at $z$=7.5. Orange-gray symbols show the mean $\Ebv$ as a function of halo mass. The color shading indicates the mean metallicity in solar units. The maximum value of $\Ebv_{\rm max}=3.3\times10^{-3}\ (A_{\rm V,max}=1.3\times10^{-2})$ is reached for $\log(M/M_{\rm halo})=10.15$, where $\log(Z/\Zsun)=-2.12$. The black dashed and dash-dotted lines represent the mean extinction for haloes with $\log(M/M_{\rm halo})\geq 8.0$ and $9.0$, respectively. The blue symbols represent mean $\Ebv$ for a given halo mass range with 1$\sigma$ error bars. A linear least-squares fit to the mean values is provided as the solid black line.
}
\label{fig:fig2}
\end{center}
\end{figure}

In Figure~\ref{fig:fig2}, we present the mean halo $\Ebv$ as a function of halo mass with the mean metallicity indicated by the colour scale.  There is a clear correlation between halo mass, extinction and the amount of metals found in the halo.  This relation is due to the fact that higher mass haloes typically experience higher star formation rates, which will lead to more metals injected into the ISM and, consequently, more dust. The relationship is well described by a power-law $\left<\Ebv\right>\propto M_{\rm halo}^{0.80}$. 
The maximum Pop~III extinction occurs in a halo with $\log(M/M_{\rm halo})=10.15$, which has a mean metallicity of $\log(Z/\Zsun)=-2.12$, leading to $\Ebv_{\rm max}=3.3\times10^{-3}\ (A_{\rm V,max}=1.3\times10^{-2})$. On average, we find that haloes with $\log(M/M_{\rm halo})\geq 8.0$ have $\left<\Ebv\right>\approx2.1\times10^{-4}$, and those with $\log(M/M_{\rm halo})\geq 9.0$ show $\left<\Ebv\right>\approx 8.7\times10^{-4}$. Averaged over the entire halo, Pop~III metals thus have a negligible impact on the intrinsic $\buv$.

\subsection{Column density}
\label{subsec:dens}
In the previous section, we averaged physical properties over the
entire halo gas. The downside of this approach is that high density
and high metallicity regions in the ISM can be diluted. To explore this possibility, we examine the properties of each gas particle, or resolution element, contained in our simulation volume. We are thus more accurately modeling the Pop~III extinction experienced by a photon if it were to pass directly through a given gas cloud. 

\begin{figure}
\begin{center}
\includegraphics[scale=0.34] {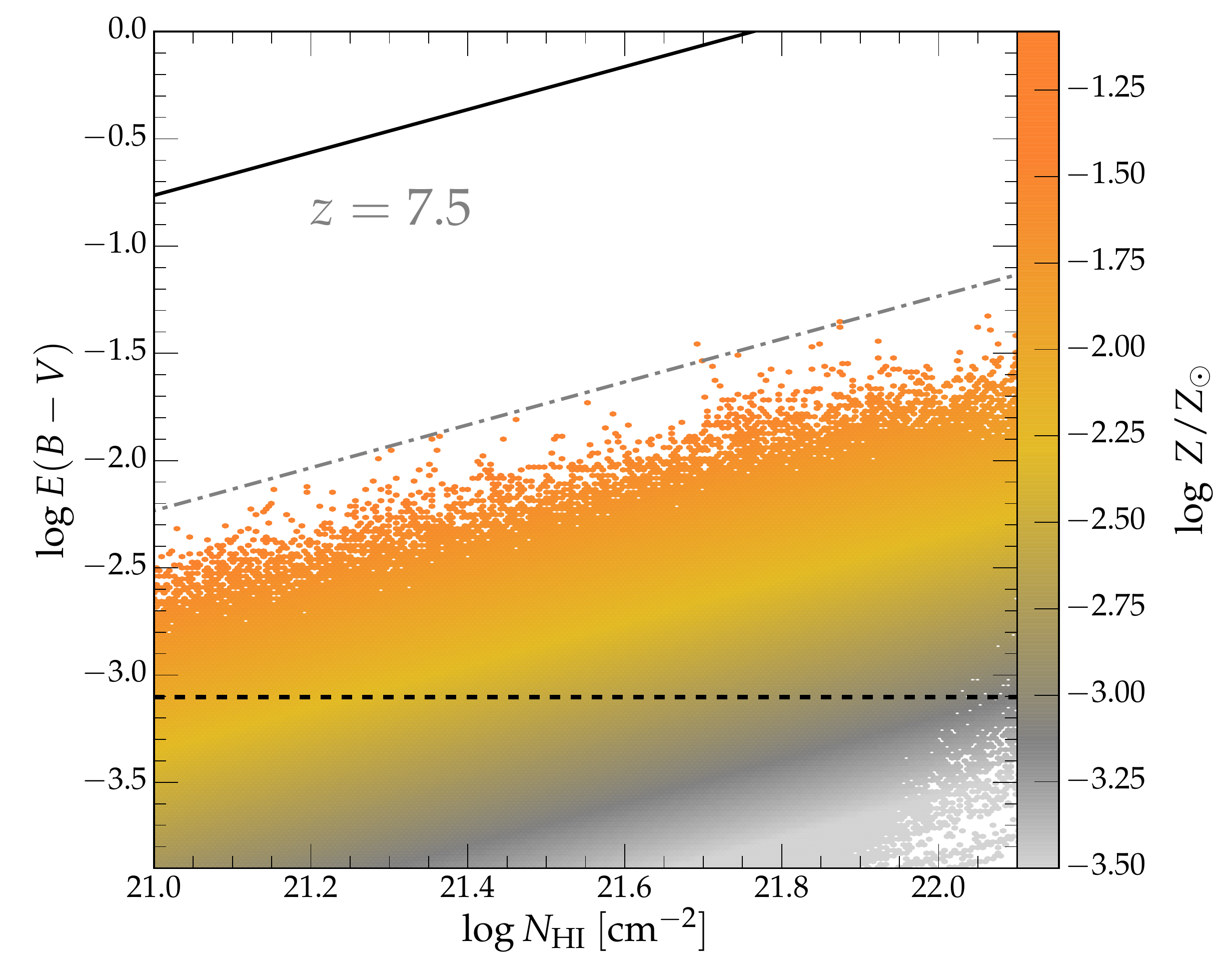}
\caption{2-D histogram of the \ion{H}{i} column density ($N_{\ion{H}{i}}$) vs. the calculated $\Ebv$ with hexagonal binning for gas bound to a dark matter halo. As in Fig.~\ref{fig:fig2}, color indicates gas metallicity, and $\Ebv$ is calculated by taking the characteristic length to be the local Jeans length ($L_{\rm char} = L_{\rm J}$). The dashed black line represents the mean Pop~III extinction ($\left<\Ebv\right>=6.6\times10^{-4})$ for all non-zero metallicity halo gas particles. For reference, the black solid line represents the Milky Way normalized value (Equ.~\ref{eq:ebv}), and the gray dashed-dotted line is a scaled down version to guide the eye.} 
\label{fig:dens_ebv}
\end{center}
\end{figure}

In Fig.~\ref{fig:dens_ebv}, we show $\Ebv$, calculated from
Equ.~\ref{eq:ebv}, versus column density, with colour corresponding to
gas metallicity, for each particle. Here, we use the local Jeans length, calculated for each gas particle, as the characteristic length scale ($L_{\rm char} = L_{\rm J}$). Since our simulations do not directly include Pop~II star formation, we would tend to over-estimate the extinction in dense, metal-rich gas using the above methodology, as feedback from Pop~II star formation would disrupt such clouds. To compensate, we exclude any gas particles which would qualify for Pop~II star formation in a standard sub-grid model (i.e. $n>100\ {\rm cm^{-3}}$ and $Z>10^{-4}\ \Zsun$).  The metals contained in such particles are distributed to the surrounding ISM so as to conserve total metal mass. 
We find a maximum Pop~III extinction of $\Ebv_{\rm max}=0.03\ (A_{\rm V,max}=0.13)$, considerably higher than the estimate from Sec.~\ref{subsec:halo}. The average for all non-zero metallicity gas particles in Fig.~\ref{fig:dens_ebv} is $\left<\Ebv\right>\approx6.6\times10^{-4}$.

\subsection{Galaxy LOS}
\label{subsec:los}
\begin{figure}
\begin{center}
\includegraphics[scale=0.34] {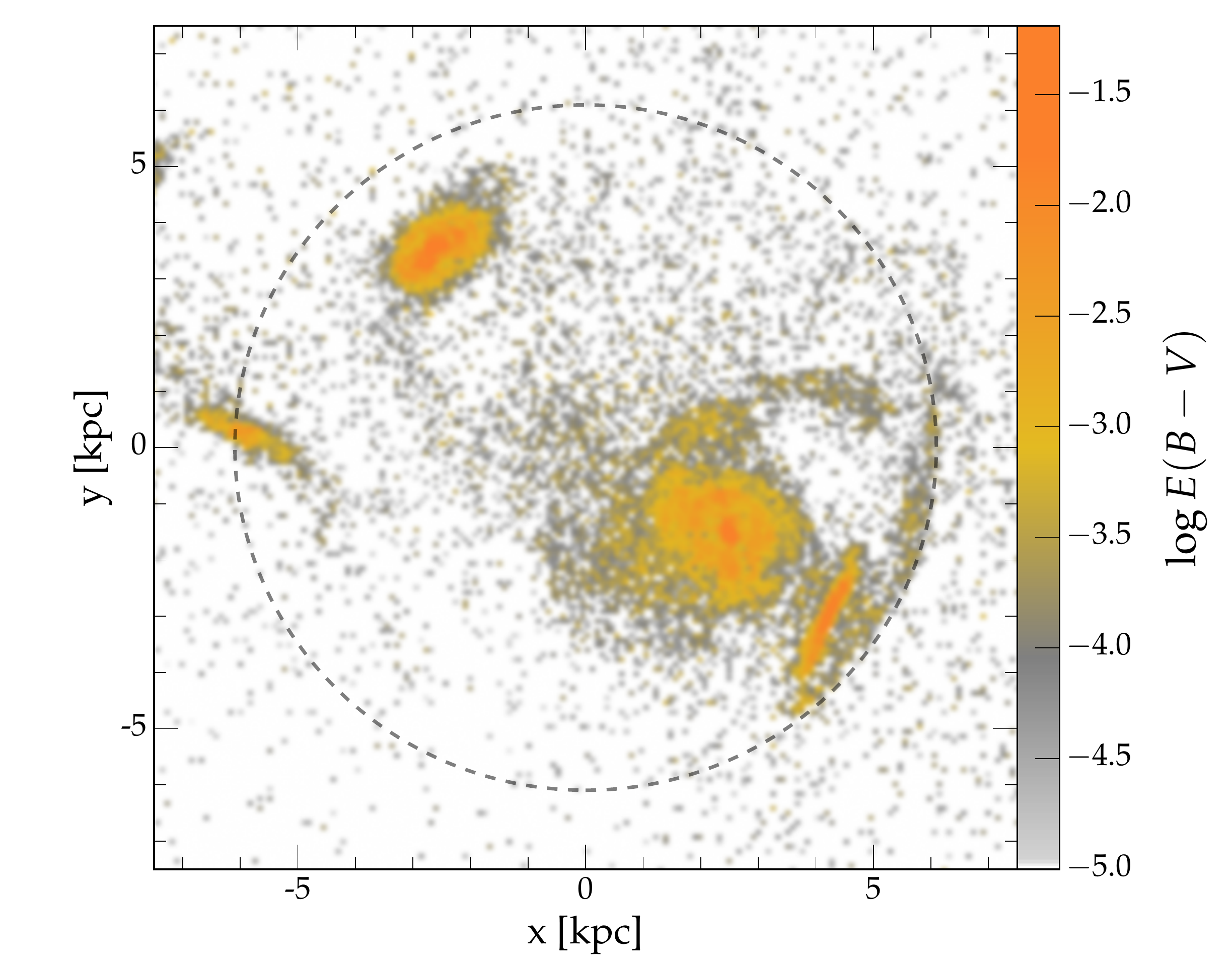}
\caption{Projected $\Ebv$, calculated for the halo with the highest simulated extinction level ($M_{\rm halo}\approx10^{10}\ \Msun$). Each pixel represents the sum along the z-axis. Along the most metal-rich, dense LOS we find $\Ebv_{\rm max}=0.07\ (A_{\rm V,max}=0.28)$, and a mean value of $\left<\Ebv\right>=3.4\times10^{-4}$.  The dashed circle indicates the half mass radius ($R_{1/2}$) for this halo. Note that the gray pixel ``noise'' is an artifact of the binning process, but does represent the overall extent of zero-metallicity gas in the image.} 
\label{fig:fig4}
\end{center}
\end{figure}

Finally, we examine individual LOS directions through the galaxy identified in Sec.~\ref{subsec:halo} to exhibit the highest average Pop~III extinction value. Here, we use a binning process to place each particle in a grid cell, after which we sum the gas and metal mass of each cell along the $z$-axis. The column density and subsequent extinction is then calculated for each LOS using Equ.~\ref{eq:ebv}, shown in Fig.~\ref{fig:fig4} with colour representing the $\Ebv$ due to Pop~III enrichment. The gray pixels indicate the overall extent of the zero-metallicity gas.
With this method we find a maximum of $E(B-V)_{\rm max}=0.07\ (A_{\rm V,max}=0.28)$, and a mean value of $\left<\Ebv\right>=3.4\times10^{-4}$, only slightly higher than those found in Sec.~\ref{subsec:dens}.  This is not unexpected as we are summing the extinction along a LOS through the most dense and metal rich region of the galaxy.  However, this maximum value should be viewed as an extreme upper limit as it represents extinction experienced though the entire LOS ($\sim 2.4$ kpc).  A more likely scenario is that star forming regions would be embedded at different depths along the LOS and thus experience varying levels of extinction.  A more realistic emerging spectrum would be composed of a composite of these spectra and likely present less attenuation. 

\subsection{Baseline Pop III spectral slope}
\label{subsec:baseline}
We now employ a Monte Carlo sampling technique to produce representative model spectra, and derive spectral slopes, $\buvpiii$, for Pop~III enriched galaxies. We randomly sample over a range of dark matter halo mass, SSP age and metallicity, $f_{\rm esc}$, and $\Ebv$.  Dust extinction is sampled from the relationship between halo mass and metallicity provided in Figure~\ref{fig:fig2}, and $f_{\rm esc}$ is taken from results presented in \citet{Paardekooper:15}, who use the FiBY simulations to derive distribution functions for $f_{\rm esc}$ as a function of halo mass.  Halo mass, age and SSP metallicity are drawn from a uniform distribution.  For each set of parameters, a model spectrum is created \citep{Schaerer:03b}, in turn providing the resulting UV spectral slope.In the case of age and intrinsic metallicity, we use linear interpolation between the two closest models.  The full list of parameters, references and range of possible values can be found in Table~\ref{tbl:params}.

\begin{table} 
\caption{Parameter ranges for Monte Carlo sampling to calculate baseline $\buvpiii$, as presented in Fig.~\ref{fig:fig5}. For $f_{\rm esc}$, we adopt results from the FiBY simulations \citep{Paardekooper:15}, which provide escape fraction probability distributions as function of halo mass. Based on results in Finkelstein 2017 (in-prep), we apply a correction factor of $\sim 5$ to the \citet{Paardekooper:15} escape fractions in order to reconcile observations and re-ionization calculations. All parameters are drawn uniformly between lower and upper limits, unless otherwise constrained by an underlying distribution, as is the case for $f_{\rm esc}$ and $\Ebv$. 
}
\begingroup
\setlength{\tabcolsep}{6pt} 
\renewcommand{\arraystretch}{1.2} 
\begin{center}
\begin{tabular}{ccc}
\hline
Parameter  	& 	Range & Source	 	\\
\hline
DM halo mass		& $10^8-10^{10}\ \Msun$		&	Simulation resolution 	\\
SSP age				& $10-50$ Myr 				&	Observed estimate 		\\
SSP metallicity & $10^{-5}-10^{-1}\ \Zsun$		&	\citet{Jaacks:17a}	\\
$f_{\rm esc}$ 	& $0-1$							&	\citet{Paardekooper:15}\\
$\Ebv$			& $\propto M_h^{0.80}$	&   This work (Figure~\ref{fig:fig2})\\	
\hline
\end{tabular} 
\label{tbl:params}
\end{center}
\endgroup
\end{table}

\begin{figure}
\begin{center}
\includegraphics[scale=0.34] {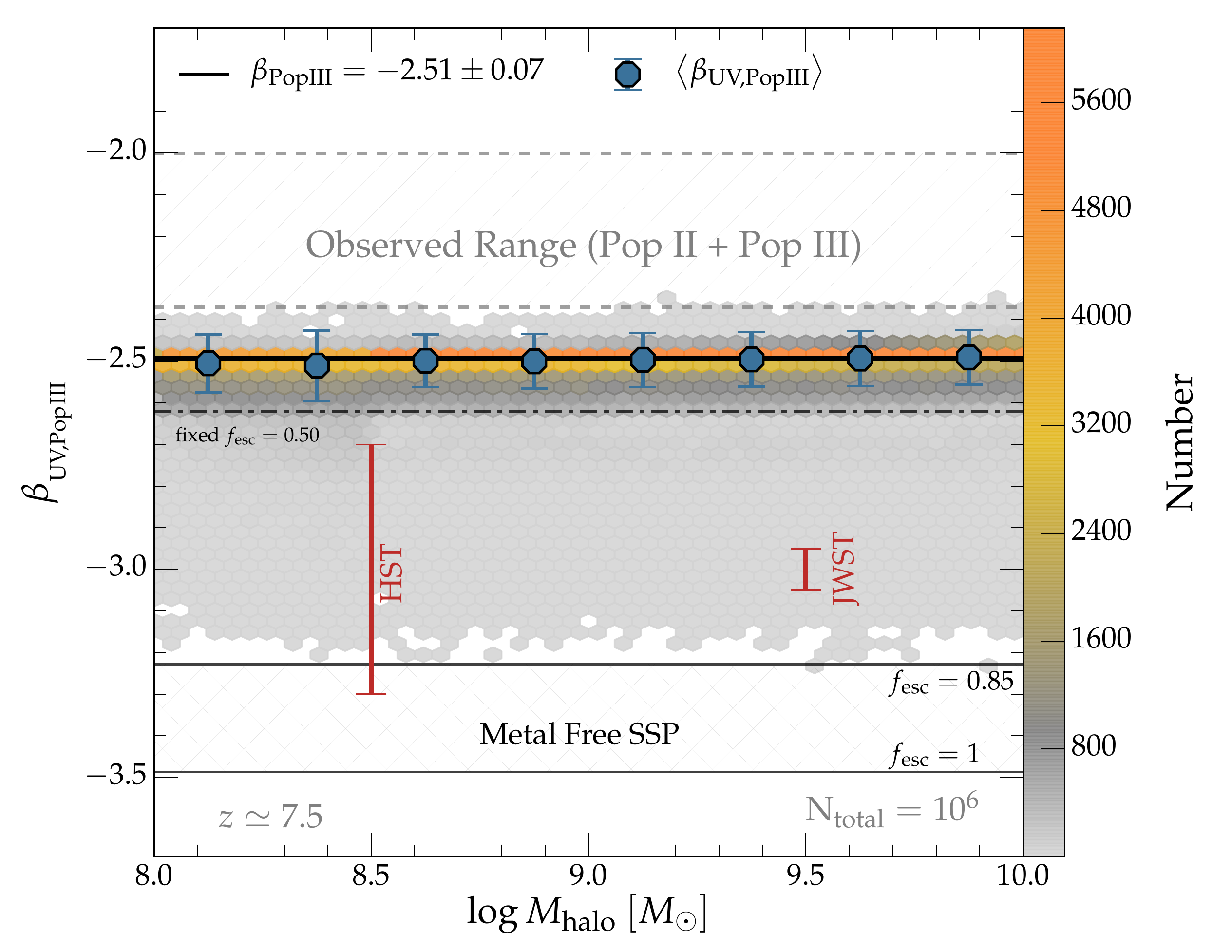}
\caption{ UV spectral slopes from Pop~II SSPs reddened by Pop~III dust, $\buvpiii$, calculated from the Monte Carlo sampling of the relevant parameter space (see Table~\ref{tbl:params}). Blue symbols indicate median values with 1$\sigma$ error bars, and the solid black line is a linear least-squares fit. The gray-orange shaded region represents the density of calculated $\buvpiii$ values. The observed range of Pop~II + Pop~III extinction at $z \sim 7$ and $M_{\rm UV}\sim -18.5$ is indicated by the region with diagonal lines (from \citealt{Finkelstein:12c,Bouwens:14a,Dunlop:13}). The red error bars denote the approximate uncertainty on measuring $\beta$ at this magnitude with the {\it HST}, and the expected improvement with {\it JWST}.
} 
\label{fig:fig5}
\end{center}
\end{figure}

In Figure~\ref{fig:fig5}, we present the results of our random sampling study. The most striking feature is that the median value of $\left<\buvpiii\right>= -2.5\pm0.07$ appears to be roughly constant across all halo masses considered in this work with only a slight dip at $M_{\rm halo}=10^{8.5}\ \Msun$.  This can be understood by further examination of the primary contributors to $\Delta\beta$, $\Ebv$ and $f_{\rm esc}$ (see bottom panels of Fig.~\ref{fig:fig1}). In the case of $\Ebv$, there is a strong correlation of increasing Pop~III extinction with increasing halo mass (Fig.~\ref{fig:fig2}).  However, the amount of Pop~III extinction is extremely low ($\Ebv<10^{-3}$), with negligible impact on the $\buvpiii$ slope, leaving $f_{\rm esc}$ as the primary factor. The mean $f_{\rm esc}$, calculated from the \citet{Paardekooper:15} probability distributions, are also roughly constant over our range of halo masses, with $\left<f_{\rm esc}\right>=0.14,0.06, 0.15, 0.15$ for $\log M_{\rm halo}/\Msun=8.0, 8.5, 9.0, 10.0$, respectively. It is worth noting that our adopted values for $f_{\rm esc}$ are very consistent with those found in \citet{Wise:14} at the upper limit of their studied mass range ($M_{\rm halo}/\Msun=8.0, 8.5$). Finally, the near redshift independence of $f_{\rm esc}$ implies a $\buvpiii$ floor which is also largely redshift independent.

Over the same age range, our primary result of $\left<\buvpiii\right>= -2.51\pm0.07$ is consistent with the study by \citet{Zackrisson:13}, who assumed a fixed SSP metallicity and $f_{\rm esc}$ to determine $\beta$ as a function of age. The primary improvement here is that SSP metallicity and $E(B-V)$ are directly extracted from our simulation, as well as the utilization of detailed distribution functions for $f_{\rm esc}$ from \citet{Paardekooper:15}. Our work is also differentiated by the fact that we focus on the most extreme, chemically primitive scenario, in and effort to predict the baseline $\beta$ value, whereas \citet{Zackrisson:13} explore a wide parameter space to offer predictions for many physical settings.

\section{Summary and Conclusions} 
\label{sec:sum}
Using cosmological volume simulations and a custom built sub-grid model for Pop~III star formation, we examine the baseline dust extinction due to metals produced by Pop~III in the first billion years of cosmic history. Our major conclusions are as follows:
\begin{itemize}
\item {Dust extinction due to Pop~III star formation is strongly correlated with halo mass when averaged over the system, expressed as $\Ebv\propto M_{\rm halo}^{0.80}$ (Fig.~\ref{fig:fig2}). However, the overall extinction due to Pop~III is very low ($\Ebv<10^{-3}$) and contributes little to changing the overall $\beta$ slope.}
\item {When considering column densities of individual gas clouds which are bound to dark matter haloes, we find an average Pop~III extinction of $\left<\Ebv\right>=6.6\times10^{-4}$.}
\item {Integrated lines-of-sight within a given halo can, in very limited cases, exhibit Pop~III extinction values as high as $\Ebv_{\rm max}=0.07\ (A_{\rm V,max}=0.28)$ along the most metal rich, dense LOS (Fig.~\ref{fig:fig4}). }
\item {Statistical Monte Carlo studies of constrained parameters which control  $\buvpiii$ (halo mass, age, metallicity, escape fraction, and dust extinction) suggest that $\left<\buvpiii\right>=-2.51\pm0.07$ would be representative of extremely metal poor galaxies with $Z<10^{-2} \Zsun$. }
\end{itemize}

In interpreting the results presented in Fig.~\ref{fig:fig5}, consider
a scenario in which a future deep-field observation detects an object
at $z\geq7.5$ with a $\buv\sim-3.0$. {\it What would this tell us?}
Our analysis here suggests that such an extreme UV spectral slope
would be indicative of being generated by a very young, Pop~II star
forming galaxy, which experiences minimal nebular emission and has
been enriched by only Pop~III-produced metals. Such systems would be
excellent targets for deep ground-based spectroscopic follow-up
observations with resources such as the Giant Magellan Telescope
(GMT). Given current error bars for $\buv$ measurements ($\pm0.3$ at
$M=-$18, the limit of the Hubble Ultra Deep Field at $z\!>$7), it
would be challenging to uniquely identify a truly Pop~III star forming
galaxy, with estimated slopes of $\beta_{\rm UV}\lesssim-3.0$. Future
$\it JWST$ deep-field campaigns will have significantly reduced error
bars at these same magnitudes ($\pm0.05$), offering the exciting possibility to uniquely identify low-$Z$ stellar populations.

Our results indicate that the reddening experienced by $z\geq7$ dwarf
galaxies is almost entirely due to nebular emission. However, we here
have considered only the baseline enrichment from Pop~III star
formation. Therefore, it is useful to provide a rough estimate of the
amount of Pop~II created metals which are missed in our Pop~III only
simulation. For this purpose, we assume a halo of $M_{\rm vir}\approx
10^{10}\ \Msun$ with $Z_{\rm PopIII}\approx10^{-2}\ \Zsun$, the
plateau Pop~III enrichment found in \citet{Jaacks:17a}. In such a
halo, we have a baryonic mass of $M_{\rm B}\approx 10^9\ \Msun$, which
results in a total mass in metals of $M_{\rm Z,PopIII}\approx10^4\
\Msun$. To derive the Pop~II metal enrichment, we use abundance
matching arguments \citep{Behroozi:13} to arrive at a corresponding
 stellar mass of $M_{\rm *,tot}\approx10^7\ \Msun$.  For a
standard IMF \citep[e.g.][]{Salpeter:55,Chabrier:03,Kroupa:01}, there
is approximately one core collapse SN (CCSN) for every $\sim100 \Msun$
($\eta_{\rm CCSN}\sim0.01$). If we further assume that a typical CCSN
has a progenitor mass of $M_*=10\ \Msun$ and a metal yield of $y_Z\sim
0.01$, we arrive at a total metal mass of $M_{Z,\rm PopII}\approx10^4\
\Msun$  (i.e. $M_Z\approx \eta_{\rm CCSN}M_{\rm *,tot}M_*y_Z$). This
estimate, which is consistent with the more sophisticated analysis by
\citet{Mancini:15}, illustrates that $M_{\rm Z,PopIII} \approx M_{\rm
  Z,PopII}$. Thus, even with Pop~II star formation included, high-$z$
dwarf galaxies are unlikely to experience a measurable degree of dust
extinction, consistent with current observations. In future work, we will revisit this estimate with Pop~II star formation self-consistently enabled in our simulations.

There are several caveats to this work relating to the uncertainty of several adopted models and parameters. In particular there is a high dependence upon the accuracy of the SSP models utilized \citep{Schaerer:03b}.  If future advancements in stellar evolution and nucleosynthesis lead to more accurate modeling of extremely low-metallicity stellar populations, then the results here should be revisited. We here also adopt stellar evolutionary models which assume a constant star formation history. While it has been shown, both observationally and theoretically, that high-redshift galaxies exhibit a stochastic, rising star formation history \citep{Papovich.etal:11,Finlator.etal:11,Jaacks.etal:12b,Shimizu:14,Zackrisson:17}, the low metallicity models utilized here pre-date these results.

We also adopt a fixed metal-to-dust ratio when calculating our extinction values.  It has been demonstrated by \citet{Mattsson:14} that this value could be dependent on metallicity, such that higher $Z$ leads to a larger dust ratio. In the end, our choice for the metal-to-dust parameter has no impact on our results.  Even if we doubled our adopted value so that 100\% of the metals were converted into dust, the metallicity from Pop~III star formation is too low ($\lesssim 10^{-3}$) to significantly affect the reddening. It has also been suggested by \citet{Schneider:16} that the dust content in high-redshift systems may not solely depend on metallicity, as ISM density conditions may affect the efficiency of dust formation. Again, given the extremely low metallicities, the efficiency of dust creation will have no impact on the results reported here.

Our analysis agrees with previous studies which indicate that, in the absence of dust, Lyman continuum (LyC) radiation, re-processed into nebular continuum emission, is the dominant reddening factor for a given halo \citep[e.g.][]{Zackrisson:13,Zackrisson:17,Wilkins:13,Barrow:17,Dunlop:12}. In this work, we adopt the results presented in \citet{Paardekooper:15}, who performed sophisticated radiative transfer calculations on $\sim 75,000$ simulated galaxies, taken from the FiBY simulations, to establish an escape fraction distribution for each halo mass studied. While the $f_{\rm esc}$ distribution is rather broad for each halo mass, our random draw of $10^6$ realizations exhibits a mean of $\left<f_{\rm esc}\right>=0.08\pm0.18$, when averaged over all halo masses. This implies that the majority of LyC photons are absorbed in the local ISM. Due to the strong dependence of our results on this parameter, and the uncertainties in ascertaining its physical nature, it is useful to consider the more extreme case of a fixed $f_{\rm esc}=0.50$. For this scenario we find $\beta\approx -2.62$, which is indicated by the black dash-dotted line in Figure~\ref{fig:fig5}. A physical justification for such high escape fraction could be a star forming event which has, due to strong stellar feedback, evacuated the surrounding gas, thus facilitating the escape of LyC photons into the low-density intergalactic medium. 

Finally, there is the uncertainty related to the Pop~III IMF. In our simulations, we stochastically sample each individual Pop~III stellar population from a top-heavy IMF, which is essentially (logarithmically) flat within the range $8\leq M/M_\odot \leq 140$. This IMF is representative of results from multiple high-resolution simulations which follow the collapse of a Pop~III star forming region to very high density \citep[e.g.][]{Greif:11,Stacy:13}. While there are differences in the details of the predicted Pop~III IMF across studies, the consensus view is that it is more top heavy compared to Pop~II. Should future work indicate that the Pop~III IMF is not as top heavy as currently predicted, such revision would be unlikely to impact our results, as the resulting overall metal production per stellar mass would remain low. This conclusion is supported by the results from \citet{Pallottini:15}, indicating that the Pop~III star formation rate density is largely unchanged, regardless of the assumed IMF. Testing our ``blue limit predictions'' with upcoming deep-field observations will provide an exciting view into the initial stages of cosmic chemical evolution. 

\section*{Acknowledgments} 
\label{sec:ack}
JJ would like to thank Daniel Schaerer for access to low metallicity
and Pop~III stellar evolution models. This work was supported by
HST-AR-14569.001-A \&  HST-AR-15028.001-A (PI Jaacks), provided by NASA through a grant from the Space Telescope Science Institute, which is operated by the Association of Universities for Research in Astronomy, Inc., under NASA contract NAS5-26555. VB is supported by NSF grant AST-1413501. JJ and SLF acknowledge support from the NASA Astrophysics and Data Analysis Program award \#NNX16AN47G issued by JPL/Caltech. This work used the Extreme Science and Engineering Discovery Environment (XSEDE), which is supported by National Science Foundation grant number ACI-1548562, allocation number TG-AST120024.

\bibliographystyle{mnras}

\bsp	
\label{lastpage}
\end{document}